**Contested Temporalities in Critical Minerals and Resource Extraction for Electric Vehicles**

Joseph Nyangon

May 2026

## Abstract

The increasing demand for critical minerals like cobalt and lithium, driven by electric vehicle (EV) manufacturing, presents a paradox between rapid industrial expansion and long-term sustainability. The concentration of extraction in regions such as the Democratic Republic of Congo (DRC), Chile, and Argentina has led to significant socio-environmental challenges, including ecosystem degradation, labour exploitation, and the displacement of Indigenous communities. In the DRC, cobalt mining is often linked to child labour and hazardous working conditions, while lithium extraction in Chile intensifies water scarcity, threatening local agriculture and biodiversity. The urgency to secure these resources, reinforced by regulatory measures like the U.S. Inflation Reduction Act (IRA), aims to promote ethical sourcing, yet the extraction-driven approach often exacerbates global inequalities. This chapter critically examines these contested temporalities, where the immediate economic incentives of resource extraction conflict with long-term environmental and social objectives. By advocating for community-centred governance, sustainable mining, and circular economy strategies—including recycling and material substitution—it proposes a place-based framework to balance resource security with equity, ensuring that the transition to EVs does not perpetuate the very injustices it seeks to mitigate.

## 1 Introduction to Contested Temporalities and Resource Extraction

The temporalities of resource extraction pose significant challenges to sustainable electrified transportation development, with potential supply bottlenecks in the short term. This is particularly relevant for strategic materials such as lithium, cobalt, and rare earth minerals. The surge in demand for critical raw materials, including rare earths, due to the growth of electric vehicles (EVs) has raised sustainability issues and the need for global governance responses. Following the U.S. Department of Energy (DOE) commissioned first report on their strategy for critical minerals due to the clean energy technology back in 2010, which defines 'criticality' of minerals as a "measure that combines their importance to the clean energy economy and risk of supply disruption…especially in the medium and long term" (US DOE, 2010). Table 1.1 and Appendix A present a list of critical minerals essential for the energy transition. The table includes their primary applications, significance, most representative countries where these minerals are sourced, and potential injustices associated with their extraction, processing, manufacturing, distribution, operation, and end-of-life stages. The issue of metal resources in lithium-ion batteries, a key component of EVs, is also a concern. However, promoting social, economic, and environmental sustainability through proactive stakeholders and sustainable practices such as ethical sourcing, recycling, and responsible mining can mitigate these challenges. Such efforts can lead to equitable and environmentally friendly electrified transportation systems that support social, economic, and environmental sustainability while contributing to achieving global climate targets.

Table 1.1 Critical minerals and rare earth elements essential for the energy transition and related activities

| Mineral | Applications | Significance | Representative Countries | Potential Injustices |
|---|---|---|---|---|
| Aluminium | Lightweight, transportation, packaging, construction | Energy-efficient, recyclable, infrastructure, low density | China, Russia, Canada, Australia, India | Environmental degradation from bauxite mining, worker exploitation, pollution from refining, land displacement |
| Cobalt | Batteries, superalloys, catalysts, electronics | Energy storage, high-performance materials, catalytic processes, technology integration | Democratic Republic of Congo, Russia, Australia, Philippines, Cuba | Child labour, conflict financing, environmental degradation, worker exploitation |
| Copper | Electrical wiring, plumbing, renewable energy, electronics | Conductivity, essential infrastructure, renewable energy systems, durable material | Chile, Peru, China, USA, Congo | Mining pollution, labour rights violations, environmental damage, water contamination |
| Graphite | Batteries, steelmaking, lubricants, refractories | Energy storage, industrial processes, lubrication, high-temperature applications | China, Mozambique, Canada, USA, India | Environmental degradation, worker exposure to graphite dust, mining conflicts, recycling issues |
| Indium | Touchscreens, solar cells, LEDs, alloys | High-tech displays, renewable energy, energy-efficient lighting, material enhancement | China, South Korea, Canada, Japan, Belgium | Mining environmental impacts, labour rights issues, resource scarcity, recycling challenges |
| Lithium | Batteries, energy storage, EVs electronics | Energy storage, renewable energy integration, mobility solutions, high demand | Australia, Chile, China, Argentina, USA | Water resource depletion, environmental degradation, labour exploitation, geopolitical tensions, biodiversity issues |
| Manganese | Steelmaking, batteries, alloy production, chemical applications | Strengthening steel, energy storage, industrial applications, chemical processes | South Africa, Australia, China, Gabon, Brazil | Environmental degradation, worker exploitation, mining conflicts, pollution |
| Nickel | Batteries, stainless steel, alloys, electronics | Energy storage, corrosion resistance, high-performance materials, technological applications | Indonesia, Philippines, Russia, Canada, Australia | Mining pollution, labour rights violations, environmental harm, resource conflicts |
| Platinum Group | Catalysts, automotive, | Emission control, high-tech devices, industrial processes, luxury goods | South Africa, Russia, | Mining environmental damage, labour rights issues, resource |

| Metals (PGMs) | electronics, jewellery | | Zimbabwe, Canada, USA | conflicts, recycling challenges |
|---|---|---|---|---|
| Rare Earth Elements | Magnets, electronics, renewable energy, defence | High-tech applications, energy transition, military technology, industrial uses | China, USA, Australia, Russia, India | Environmental pollution, labour exploitation, geopolitical tensions, resource conflicts |

Sources: Chen et al., 2024; Nyangon & Darekar, 2024; Agusdinata & Liu, 2023; IEA, 2023; IMF, 2023; Dall-Orsoletta et al., 2022; Nyangon & Byrne, 2021; Valero et al., 2021c; Habib et al., 2020; Parra et al., 2020; Banza Lubaba Nkulu et al., 2018; Nyangon et al., 2017

Achieving sustainable mineral extraction is essential for developing equitable and environmentally friendly electrified transportation systems, ensuring a just energy transition. Sovacool & Dworkin (2014) define a just energy system as "a global energy system that fairly disseminates both the benefits and costs of energy services and one that has representative and impartial energy decision-making." Addressing challenges in value chain security, strategic capacity, sustainability, and environmental justice necessitates collaboration among stakeholders. Innovative solutions can mitigate contested temporalities in resource extraction while promoting economic and environmental sustainability. Ethical sourcing practices, responsible mining, and sustainable recycling of raw materials are crucial for minimizing negative impacts and supporting sustainable EV production. However, commodity market fragmentation could increase decarbonization costs. Essential minerals such as copper, nickel, cobalt, and lithium are key inputs for EVs, batteries, and renewable-energy technologies, including solar panels and wind turbines. Demand for these minerals is projected to surge (IEA, 2023), potentially making them as economically significant in a net-zero-emissions scenario as crude oil (Boer et al., 2024).

Under a net-zero-emissions scenario by 2050, the International Energy Agency (IEA) anticipates copper demand to grow 1.5 times, nickel and cobalt demand to double, and lithium demand to rise sixfold by 2030 (IEA, 2023) (Figure 1.1). This sharp increase could drive significant price hikes, given the challenges in scaling up mining and refining operations, which remain highly geographically concentrated. Chile and Peru together mine more than a third of the world's copper, while Indonesia and the Philippines account for about half of global nickel production. Commodity markets are crucial channels through which geopolitical fragmentation impacts the global economy (IMF, 2023). The collision between geopolitical fragmentation, tensions, and interdependencies introduces frequent price volatility and global supply disruptions and risks, expected to become more common in the decade ahead (Nakanwagi, 2024). The strategic use of critical materials in geopolitical competition—particularly in clean energy and advanced electronics—intensifies economic and diplomatic bloc formations, with China and the West at the centre of this rivalry. Furthermore, demand and supply for critical minerals are geographically bifurcated: demand is concentrated in affluent populations of the Global North, while supply remains primarily sourced from the Global South (Lazard, 2023; Agusdinata et al., 2022). Trade restrictions surged in 2022, widening price differentials across regions, while foreign direct investment (FDI) in commodity sectors has been declining—a trend predating Russia's war in Ukraine. More than three years after the global economy experienced its most severe shock in 75 years, recovery remains uneven, constrained by the lingering effects of the COVID-19 pandemic, the war in Ukraine, and growing geoeconomic fragmentation.

Resource extraction temporalities complicate efforts toward sustainable electrified transportation, particularly for materials vital to EV batteries, such as lithium and cobalt. The rapid surge in global demand, largely driven by the EV industry, intensifies challenges in supply chain security, sustainability, and value chain inequities. Countries at the forefront of cobalt and lithium mining, including the Democratic Republic of the Congo (DRC), Argentina, and Chile, face not only immediate extraction pressures but also long-term social and environmental consequences. Green transition studies, particularly from Latin American perspectives, highlight the risks of "green extractivism," where large-scale resource extraction is framed as sustainable but ultimately undermines environmental and social objectives, resulting in degradation and injustice (McNelly & Franz, 2024; Ciftci & Lemaire, 2023; Valero et al., 2021c; Martinez-Alier & Walter, 2016). A collaborative, community-led approach that prioritizes environmental justice and local autonomy over corporate profits and investor security is essential (Byrne et al., 2022; Sovacool et al., 2021). Integrating ethical sourcing, responsible mining, and robust recycling practices will be crucial in establishing a sustainable EV sector aligned with global climate targets. Additionally, a green corridor agreement could play a pivotal role in safeguarding the international flow of critical minerals necessary for the green transition. Such agreements, guided by shared climate goals, can help stabilize supply chains and mitigate risks associated with geopolitical fragmentation.

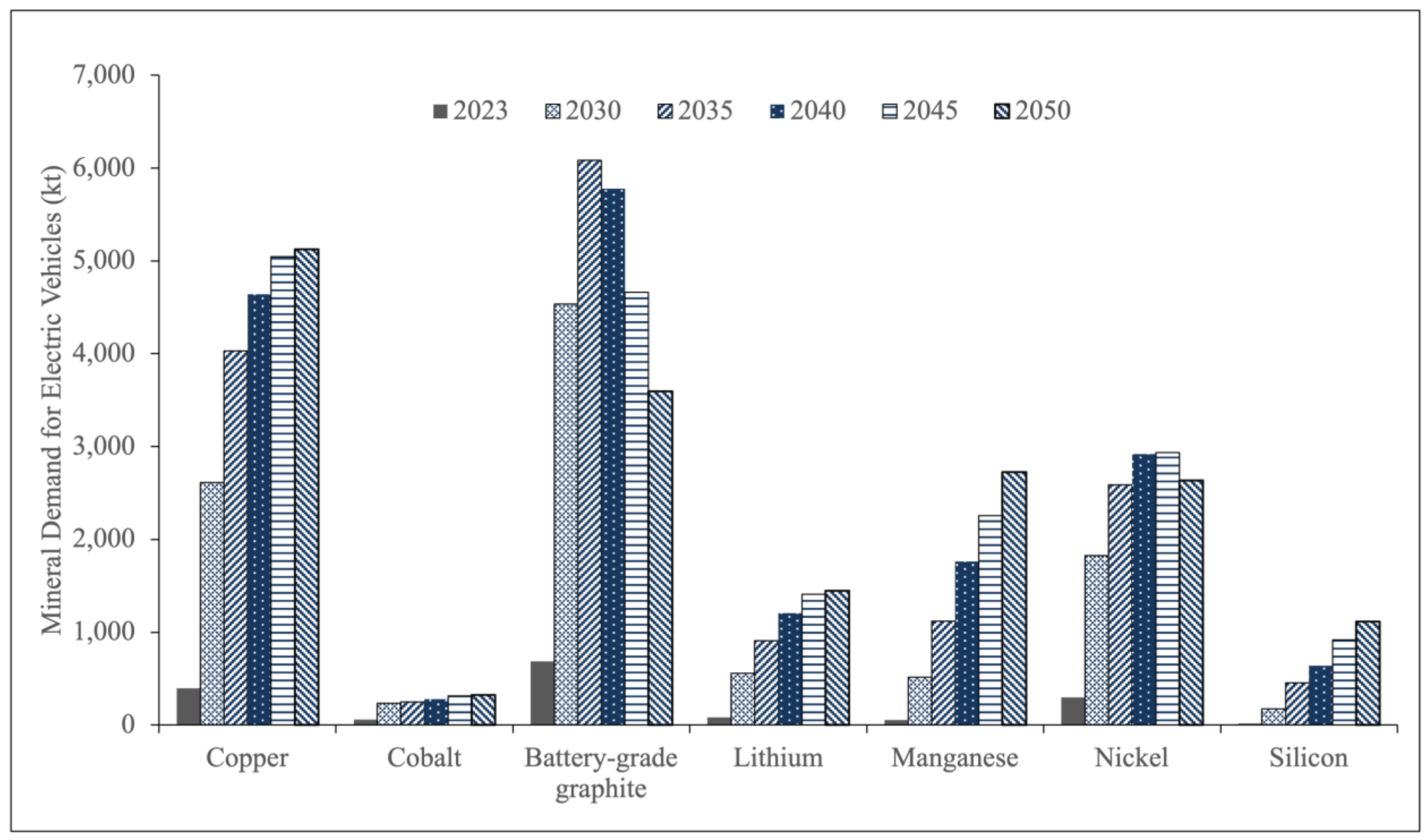

Figure 1.1 Mineral demand for EVs (kt) in a net-zero-emissions scenario. Data source: (IEA, 2023).

## 1.1 Understanding Contested Temporalities

The concept of contested temporalities in resource extraction for EVs involves conflicting perspectives on time, memory, and future expectations. Extractive industries and governments often promote accelerated development, while local communities may resist based on historical experiences and anticipated impacts (Fent & Kojola, 2020). The EV industry creates "points of continuity" with gasoline vehicles to increase adoption, navigating material constraints of electric power (Krishnan & Butt, 2022). Fictional expectations about lithium's future economic potential drive conflicts across spatial scales (Carrasco et al., 2023). The global EV transition raises justice concerns, particularly regarding resource extraction in the Global South (Agusdinata et al., 2022; Prause & Dietz, 2022). These issues challenge the notion of EVs as a purely "green" solution, potentially perpetuating global inequalities (Anlauf, 2017). Resource-rich countries like Indonesia are leveraging the energy transition to promote green development and resource nationalism (Wijaya & Sinclair, 2024).

Contested temporalities captures the conflict between immediate economic gains from resource extraction and the long-term social and environmental sustainability goals central to the low-carbon transition. This temporal dissonance is especially evident in the mining of critical materials like cobalt and lithium, essential to EV battery production, yet fraught with challenges. As nations ramp up efforts to reduce carbon emissions, the demand for these minerals has outpaced sustainable extraction rates, often resulting in environmental degradation and exploitation in resource-rich regions like DRC and Chile. The urgency to secure these resources aligns with industrial and climate ambitions yet clashes with the ecological needs of these areas. Rapid extraction can degrade ecosystems, jeopardize water resources, and disrupt local economies dependent on traditional land use practices, indicating that without effective governance, particularly polycentric authorities that contribute to informed place-based governance systems (Byrne et al., 2022; Nyangon, 2021; Ostrom, 2012), this disjunction between short-term extraction goals and enduring environmental needs undermines the very sustainability objectives that EVs seek to advance (Matanzima & Loginova, 2024).

## 1.2 The Role of Energy Transition Minerals in EV Manufacturing

The rapid adoption of EVs is crucial for reducing transportation emissions, but it has led to increased demand for critical metals like lithium, cobalt, nickel, and rare earth elements (Habib et al., 2020; Zhang et al., 2023). This surge in demand poses potential supply risks and sustainability challenges (Agusdinata et al., 2022; Jones et al., 2020). By 2030, demand for cobalt and lithium is expected to increase 37-fold and 18-fold, respectively, compared to 2015 levels (Jones et al., 2020). The limited availability of these metals may constrain large-scale EV deployment (Ballinger et al., 2019). To address these challenges, the industry is focusing on reducing cobalt content in batteries, improving recycling methods, and exploring alternative sources like seawater lithium (Luong et al., 2022; Mayyas et al., 2023). However, policy coordination and sustainable mining practices are needed to ensure a just and efficient supply chain for these critical materials (Sovacool et al., 2020).

The geographical concentration of these energy transition minerals further accentuates temporal disparities in resource extraction (Matanzima & Loginova, 2024). Countries such as the DRC, Chile, and Argentina are central suppliers of cobalt and lithium, but the wealth these resources generate rarely benefits local populations. In these regions, the heavy influence of multinational corporations often translates into socio-economic inequities, exploitative labour practices, and environmental injustices that disproportionately affect vulnerable communities. For instance, child labour and unsafe conditions plague the cobalt mines of the DRC, while the water-intensive lithium mining in Chile depletes local water sources critical to Indigenous communities and agriculture (Dua et al., 2024). Such socio-economic issues underscore the need for policies that not only address the environmental impacts of mining but also promote fair labour practices, community engagement, and equitable economic gains. As the demand for EVs continues to escalate, the extraction-driven approach exacerbates these inequalities, pushing resource-dependent communities to bear the weight of the transition without realizing its potential benefits.

To navigate the conflicting timescales of resource extraction, an integrated approach that balances immediate demands with future sustainability is imperative. Government policies, like the U.S. Inflation Reduction Act, aim to embed ethical sourcing and environmental protections within the supply chain, setting a benchmark for responsible resource management. However, a global cooperative framework is essential to address these contested temporalities effectively, aligning regulatory standards across borders to mitigate exploitation and environmental degradation. Furthermore, investments in research for alternative materials and recycling innovations could reduce dependency on high-impact minerals like cobalt and lithium, making sustainable resource management feasible over time. As emerging policies and innovative solutions converge, they offer pathways to a more resilient and equitable EV supply chain, one that values long-term ecological integrity and social equity alongside the technological advancements driving the electrification of transportation (Sheldon & Dua, 2024).

## 2 The Environmental and Social Challenges of Resource Extraction

The extraction of critical raw materials poses significant environmental challenges, including land degradation, water depletion, and ecosystem impacts (Manhart et al., 2019; Wall & Pell, 2017). Life cycle assessment (LCA) can help evaluate these impacts, though it has limitations in assessing ecosystem degradation and water resource effects (Mancini et al., 2015; Manhart et al., 2019). The circular economy approach and resource efficiency are crucial for addressing raw material depletion and environmental concerns (Diemer et al., 2018; Martins & Castro, 2020). Exploiting extractive waste facilities can be a sustainable alternative to primary mining, but requires consideration of economic, environmental, and social factors (Dino et al., 2020). LCA can support the management of critical raw materials by providing information on resource flows and potential supply risks (Cimprich et al., 2019; Mancini et al., 2015). Integrating criticality assessments with LCA can offer a more comprehensive understanding of raw material impacts and support decision-making for sustainable resource management (Vadenbo et al., 2014). The environmental and social challenges of resource extraction in the global quest for EV batteries are complex and multilayered, particularly in regions like the DRC and the Lithium Triangle in South America. The ecological footprint of extractive industries in these areas is both profound and far-reaching. In the Lithium Triangle—spanning Chile, Argentina, and Bolivia—lithium extraction requires vast quantities of water, a resource already scarce in the region's arid landscapes. This intense water consumption has severe downstream effects on local agriculture and biodiversity, disrupting delicate ecosystems and jeopardizing local food production (Johnson et al., 2024). For example, the saline groundwater used in lithium brine extraction directly competes with traditional agricultural water sources, drying up lagoons and wetlands essential for wildlife and plant species. In Chile's Atacama Desert, one of the driest places on Earth, this imbalance is especially pronounced, leading to land degradation and escalating tension between mining companies and local communities dependent on these same water resources for their livelihoods.

The intersection of transportation electrification and environmental justice exposes intricate equity challenges, as electrification, while reducing emissions and improving air quality, can also exacerbate disparities in air quality, health, technology access, and infrastructure development (Hennessy & Azevedo, 2024). Disparities in EV adoption, charging infrastructure placement, and economic impacts particularly affect Black, Indigenous, and People of Color (BIPOC), low-income, and frontline communities, underscoring the need for targeted policies to ensure equitable outcomes (Tsukiji et al., 2023). Evaluating energy justice and equity impacts requires addressing dimensions such as aggravated energy burdens, inaccessibility of low-carbon technologies, displacement due to renewable energy siting, and unequal health and employment benefits from fossil fuel retirement (Kime et al., 2023). The transition to e-mobility risks creating inter- and intra-regional inequalities, necessitating a critical approach to policy development. Global perspectives on just mobility futures must address these injustices at the upstream end of global EV production (Prause & Dietz, 2022). Additionally, environmental justice research highlights how extractive industries often reinforce existing social inequities (Malin et al., 2019), with resource extraction conflicts highlighting tensions in territorial control, equity, and governance (Bebbington & Bebbington, 2010).

Socially, the consequences of extraction are equally stark, with environmental justice research showing that extractive industries reinforce existing social inequities like exploitative labour practices, indigenous rights and resource access, human rights abuses, mineral procurement challenges, child labour and environmental degradation (Boateng & Klopp, 2024; Agusdinata & Liu, 2023; Malin et al., 2019), with conflicts over resource extraction highlighting tensions between territorial control, equity, and governance (Dall-Orsoletta et al., 2022; D. Bebbington & Bebbington, 2010). These social inequities can be amplified or reinforced by environmental inequities such as exposure to systemic and structural toxic hazards (Pellow, 2017). In the DRC, the cobalt mining sector, crucial for EV battery production, has been rife with abuses, including child labour and conditions often described as "modern slavery." Studies reveal that miners, many of whom are children, work under dangerous conditions with minimal safety measures or basic labour rights, exposing them to toxic dust and hazardous equipment (Sändig et al., 2024). These exploitative practices disproportionately affect women and children, who occupy some of the lowest-paying and most hazardous roles within the industry. The absence of gender equity and worker protections further entrenches poverty cycles, creating an environment where local communities are forced to trade safety and dignity for survival. The power asymmetries between multinational corporations and local labourers exacerbate this vulnerability, allowing extractive operations to continue largely unimpeded by social responsibility standards and worker protections.

The clash between the immediate demands for mineral resources and the imperative for long-term sustainability is a critical concern for the future of EV-related extraction. As the global demand for EVs drives the need for rapid mineral extraction, there is an urgent push to accelerate resource use in regions with minimal regulatory oversight. However, this focus on meeting short-term supply goals often undermines the stability and resilience of local communities, which are left to cope with the resulting environmental damage long after extraction ceases. Mining sites frequently remain contaminated, with communities facing enduring impacts from deforestation, soil erosion, and polluted water sources. These lingering environmental effects erode the long-term viability of the land, leaving behind "sacrifice zones" where once-thriving ecosystems and communities have been permanently disrupted. In this context, the green transition itself is fraught with contradictions, as efforts to support sustainable technologies inadvertently compromise the sustainability of the regions that supply their essential raw materials.

Addressing these risks require a balanced framework that prioritizes the health, well-being, and rights of affected communities alongside industrial and technological progress. Rather than emphasizing extraction at any cost, it is essential to design a global ‘energy justice’ system ‘that fairly disseminates both the benefits and costs of energy services, and one that has representative and impartial energy decision-making’(Sovacool & Dworkin, 2015). Such an approach emphasizes the complex and evolving intersection of place-based operationalization of energy justice theory across four primary axes: *recognition, procedural, distributional, and restorative* components. Recognition justice in mineral extraction for EV production highlights the need to respect the rights, values, and voices of affected communities, addressing discrimination, exclusion, and power imbalances while fostering dialogue and equitable decision-making (Nakanwagi, 2024; Zhou & Brown, 2024; Agusdinata et al., 2023). Meanwhile, procedural justice assumes inclusive, transparent decision-making mechanisms by emphasizing broad participation, voice consideration, and co-developed procedures (Rosner et al., 2023). Key aspects include unbiased frameworks, public consultation, conflict resolution, and stakeholder dialogue, addressing issues like environmental impact, social licensing to operate, and cross-border actions in critical minerals (Heffron, 2020; Lee & Byrne, 2019). Distributional justice implies equitable benefit and burden sharing, addressing disproportionate impacts on marginalized Global South communities, such as indigenous rights and resource access, especially during resource extraction (Agusdinata & Liu, 2023), while restorative justice emphasizes repairing harms and addressing historical injustices from mineral extraction, advocating environmental restoration and social compensation through legal standards (Wallsgrove, 2022; Heffron, 2020). Particularly relevant to critical mineral development, restorative justice seeks to address epistemic injustices by fostering participatory dialogue and knowledge sharing (Prause & Dietz, 2022).

Dall-Orsoletta et al. (2022) analysed the significance of these four energy justice tenets alongside the whole systems approach, which includes cosmopolitan and flexibility justice, supporting a cradle-to-grave perspective to identify the true impacts of energy projects through system interactions, noting that “without this systemic view, the impacts of extracting rare metals for manufacturing energy technologies, or biofuel production on biodiversity and land use, for example, cannot be wholly understood and, therefore, mitigated.” Figure 1.2 illustrates an integrated tenet-based and whole systems approaches to energy justice in the context of EV production. This combination reframes energy decisions as justice and ethical concerns, advocating a responsible mining paradigm that safeguards the environment and upholds the rights of local communities. Key aspects of restorative justice—such as establishing specific anti-dumping laws and standards mandating recollection of used lithium-ion batteries by manufacturers, environmental safeguards, and inclusive social compensation policies through local services and job opportunities—are critical, particularly at the resource extraction stage. Governments, corporations, and consumers in the global supply chain share the responsibility of ensuring that the benefits of electrification do not come at the expense of marginalized communities. Achieving a balance between the demands of EV production and the need for environmental and social equity requires comprehensive and collaborative efforts across all stakeholders.

If ‘just transition’ (Prause & Dietz, 2022; Zhou & Brown, 2024) is to transcend mere rhetoric, the extraction of critical minerals for EV production must be reframed within a comprehensive energy justice framework that integrates distributive, procedural, recognition-based, cosmopolitan, restorative, and flexibility tenets. Distributive justice ensures that the benefits and burdens of mineral extraction are equitably shared among all stakeholders, while procedural justice calls for inclusive decision-making processes that empower local communities in shaping mining practices (Lee & Byrne, 2019).

Recognition-based justice further mandates that the intrinsic rights and cultural identities of affected populations be acknowledged and safeguarded (Heffron, 2020). A cosmopolitan perspective extends ethical considerations beyond national borders, fostering global solidarity and accountability, and restorative justice compels the implementation of measures to redress historical and ongoing harms (Sokołowski et al., 2023; Wallsgrove, 2022). Finally, the flexibility tenet underlines the need for adaptive policies that respond to emerging challenges and technological advancements (Dall-Orsoletta et al., 2022). Together, these interlocking principles offer a holistic pathway toward sustainable resource extraction, ensuring that the development of EV production does not perpetuate social and environmental inequities but instead promotes a just, resilient future for all communities involved. Although the six tenets of the energy justice framework are interdependent and jointly applicable in remedying global disparities in just EV development, recognition justice and restorative justice are pivotal in establishing a minimum acceptable standard for universal energy access (Hazrati & Heffron, 2021).

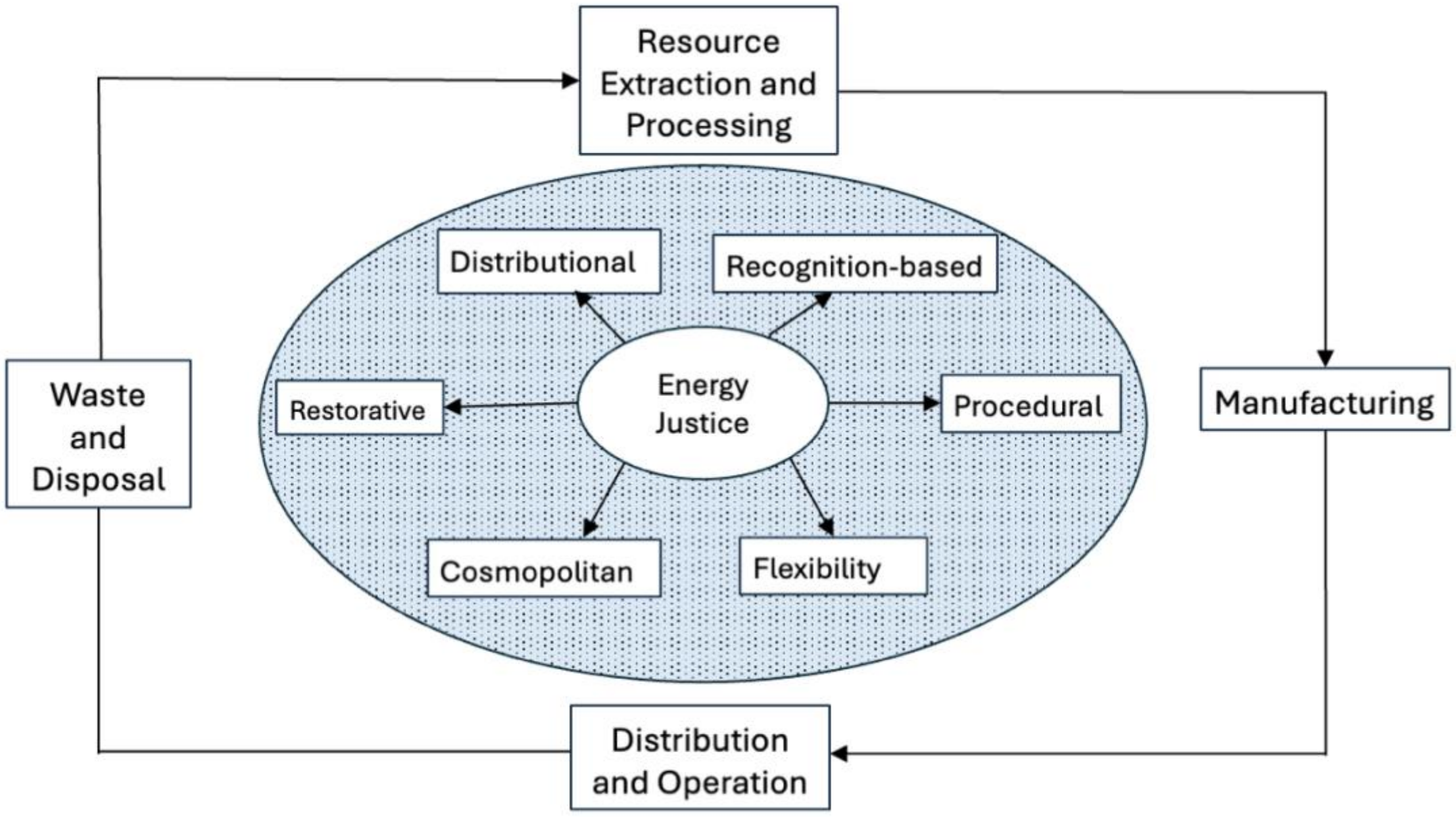

Figure 1.2 The tenet-based and the whole systems approaches to energy justice. Adapted from (Dall-Orsoletta et al., 2022)

## 2.1 The Role of Governments and Corporations in Electrification

The rapid expansion of the EV market, driven by substantial investments from both governments and corporations, reflects an unprecedented commitment to transitioning toward greener modes of transportation. This push has not only reshaped consumer landscapes but has catalysed major infrastructural developments, with governments worldwide pouring resources into EV charging networks and incentives to encourage adoption. However, this investment comes with a set of pressing ethical and environmental challenges, as the accelerated demand for critical raw materials like cobalt and lithium exerts immense pressure on extractive regions, notably in the DRC and Ethiopia. For instance in Ethiopia, socio-political and environmental impacts of the MIDROC Laga-Dambi Gold Mine, the country's largest private gold mining company, in the Oromia region has exacerbated exclusionary practices of MIDROC, which marginalized local communities, offered minimal social investment, and engaged in environmentally harmful activities (Regassa, 2022). Protests against the mine escalated in May 2018, reflecting long-standing grievances over health risks and socio-economic harms inflicted by toxic waste and heavy metal contamination. These concerns gained momentum amid Ethiopia's broader political shift toward reforms under Prime Minister Abiy Ahmed Ali.

In these areas, the interaction between corporate interests and state-backed initiatives often leads to the creation of "resource enclaves" — isolated zones where extraction occurs largely unchecked by regulatory oversight, accountability, or transparency. These enclaves allow corporations to circumvent social and environmental safeguards, resulting in significant degradation of local ecosystems and exacerbating social inequities. As a result, state-backed mining frontiers like MIDROC's can transform local landscapes into "enclaves" of extraction, undermining accountability and disenfranchising affected communities. Ultimately, it underscores how corporate-state alliances can reshape state-society relations, often at the expense of environmental justice and equitable development. As these regions struggle with compromised environmental health and labour rights, the toll of EV expansion on frontline communities becomes an unavoidable concern for all stakeholders.

## 2.2 Impact of Inflation Reduction Act on Energy and Extractive Industries

Regulatory policies play a crucial role in addressing the complexities of sustainable development, with countries like the United States advancing ethical sourcing through measures such as the U.S. Inflation Reduction Act (IRA). Passed in 2022, the IRA is a landmark policy designed to cut greenhouse gas emissions by 43-48% below 2005 levels by 2035 (Bistline et al., 2023). The IRA makes green and blue hydrogen cost-competitive with grey hydrogen (Nyangon & Darekar, 2024) and sets ambitious targets for EV battery mineral sourcing (Trost & Dunn, 2023). It promotes clean energy expansion, EVs, and domestic supply chains while mandating that companies prioritize human rights and environmental protections in their operations (von Loesecke & Chermak, 2023). By incentivizing higher labour standards and minimizing environmental impact, the IRA seeks to infuse ethical practices into the growing green economy. However, while the IRA provides a forward-looking framework, its effectiveness is limited by the global nature of supply chains. Many essential raw materials, especially those for EV batteries, come from regions with less robust regulatory oversight, where governments may lack the resources or political will to enforce strict labour and environmental protections. This disparity highlights the need for international collaboration, as unilateral policies like the IRA cannot fully address the complex, cross-border challenges inherent in EV resource extraction. The IRA's requirements for critical minerals in EV batteries pose challenges for some cathode chemistries and may benefit countries like Mexico (Hong, 2023; Trost & Dunn, 2023). Although the policy is expected to positively impact the U.S. economy, including reduced electricity costs, it may still encounter obstacles due to broader economic conditions (Bistline et al., 2023). Additionally, the IRA is reshaping global supply chains, and has stirred debate with the EU over trade and industrial policy (Bown, 2023). Meeting the heightened demand for critical minerals will require tackling supply chain, environmental, and social issues in the mining sector (Jowitt & McNulty, 2021; Sovacool et al., 2020).

A more complex dilemma arises as governments and corporations attempt to balance economic growth with the imperatives of sustainability. Although the EV industry promises substantial economic benefits, the rush to secure mineral supplies often leads to cost-cutting measures that sideline worker protections and environmental standards. In the DRC, for example, efforts by multinational companies to formalize artisanal mining have proven to be double-edged; while they introduce certain organizational structures, they frequently overlook basic worker rights. Artisanal miners—who form the backbone of cobalt extraction in the region—remain vulnerable to exploitation, working in hazardous conditions without access to adequate safety equipment, fair wages, or legal protections. This situation reveals an inherent tension in the global pursuit of sustainable technology, as the profit-oriented strategies of large corporations frequently run counter to the ethical responsibilities required to achieve true sustainability. Without comprehensive frameworks that mandate corporate accountability, economic growth in the EV sector risks becoming a vehicle for environmental harm and social inequity.

The role of corporations in this landscape is both powerful and contested. Many leading EV manufacturers and tech giants depend on complex, layered supply chains that often obscure the origins of their materials, making it challenging to ensure ethical sourcing. As consumer awareness grows, however, corporations face increased pressure to demonstrate that their supply chains are free from exploitative labour and environmental damage. Some companies have responded by setting up programs aimed at transparency, tracing their supply chains back to the mines, and even developing technology to track minerals. Despite these efforts, the systemic nature of the issues within extractive industries poses persistent obstacles. The entrenched practices of cost-cutting in resource-rich but economically fragile regions mean that voluntary corporate policies alone are insufficient to drive the deep structural changes needed. The onus thus falls on these corporations not just to implement ethical sourcing practices but to support regulatory advancements and economic initiatives that uplift local communities and protect their environments.

Moving forward, an effective strategy for sustainable electrification must bring together governments, corporations, and local stakeholders to develop an ethical, economically viable, and environmentally responsible supply chain. Collaborative international frameworks are essential for aligning policies across borders, harmonizing regulatory standards, and ensuring that ethical commitments, like those in the United States' IRA, lead to tangible global progress. The IRA, for instance, is projected to reduce retail electricity rates and have a modest macroeconomic impact, enhancing energy affordability, security, and system efficiency while promoting decarbonization and environmental stewardship (Nyangon & Akintunde, 2024; Bistline et al., 2023; Nyangon & Byrne, 2023). Furthermore, supporting community-driven initiatives within extractive regions can empower local populations, providing them with resources and opportunities beyond mining. By investing in community infrastructure and alternative economic activities, these initiatives can foster resilience and long-term economic development, allowing communities to benefit from the global demand for electrification. Integrating these efforts into the supply chain not only advances sustainable development goals but also builds trust and transparency between industry stakeholders and local communities, helping to address the social costs often associated with mining activities. As the push for electrification accelerates, a commitment to transparency, worker protections, and ecosystem conservation will be vital to ensuring that the advantages of green technologies reach beyond urban areas and affluent consumers. A truly sustainable future requires that the benefits of electrification also support the well-being of communities providing the foundational materials for this transition. By upholding ethical and environmental standards throughout the supply chain, stakeholders can help balance the promise of electrification with the broader responsibilities of social equity and environmental stewardship.

## 2.3 Impacts of IRA on Supply Chains for Electric Vehicles

The IRA legislation has fundamentally transformed the U.S. supply chain landscape for EVs by introducing impactful measures aimed at boosting adoption and domestic production. Key among these are substantial tax credits, including up to $7,500 for new EVs and $4,000 for used EVs under specific price thresholds, strategically targeted to support middle-income households through income restrictions and vehicle price caps. The IRA also imposes progressive requirements on the domestic sourcing of battery minerals and components, compelling manufacturers to establish operations in the U.S. or rely on trusted trade partners, thereby reducing dependency on foreign sources and strengthening domestic manufacturing. Complementing these efforts, production tax credits and grants extend beyond passenger vehicles to include commercial EVs and charging infrastructure, fostering a comprehensive ecosystem for EV growth. Notably, these provisions include the expansion of charging networks, particularly in rural and underserved areas, alongside environmental justice initiatives that prioritize investments in communities disproportionately affected by pollution. While the potential impact of a second-term Trump administration could introduce challenges, such as revisiting IRA-linked mandates or limiting federal funding, full repeal is unlikely due to procedural hurdles and bipartisan investments in red states. Table 1.2 illustrates the varying degrees of potential impact on IRA spending. By addressing every stage of the EV lifecycle—from production to operation—the IRA strengthens U.S. supply chain resilience, advances sustainable job growth, and promotes technological innovation across the automotive and energy sectors, creating a robust framework for a cleaner, more competitive future.

Table 1.2 Potential Trump Administration Impacts on IRA Spending and Policies

| Easy action (minimal barriers, quick implementation) | Medium action (moderate barriers, requires strategic moves) | Difficult action (high barriers, legal or political resistance likely) |
|---|---|---|
| *Temporarily delaying funds through administrative slowness:* Agencies can slow down the disbursement of IRA funds by extending review processes or adding bureaucratic hurdles. | *Reinterpret rules for tax credits and subsidies:* Modify eligibility criteria for EV mandates to favour specific industries or technologies (e.g., natural gas over EV credits or biofuels over wind/solar). | *Repeal or amend portions of the IRA*: Push for legislative changes to remove or significantly alter key provisions of the IRA (e.g., those governing EV tax credits and related incentives). |
| *Revisiting programmatic guidance documents:* Issue new or revised guidance for implementing IRA programs that are not yet implemented in favour of alternative interpretations or emphasize less aggressive climate goals. | *Scale back enforcement of climate-related goals:* Reduce the enforcement of regulations linked to IRA climate objectives, potentially leading to weaker implementation. | *Block new appropriations for IRA-related programs:* Prevent additional funding for IRA initiatives through budget negotiations, risking government shutdowns or backlash. |
| *Adjusting budget allocations within discretionary limits:* Limit approvals for loans issued through programs like the DOE's Loan Programs Office or the EPA's Greenhouse Gas Reduction Fund (GGRF). | *Altering procurement preferences:* Adjust federal procurement policies to deprioritize clean energy or climate-resilient products, affecting market demand for IRA-supported technologies. | *Litigation to challenge IRA authority:* Support or initiate lawsuits challenging the constitutionality or scope of IRA provisions, potentially leading to judicial review or reopening federal rulemaking processes to update these regulations. |
| *Public messaging against certain provisions:* Use public speeches or reports to question the effectiveness of specific IRA provisions thus delaying implementation of specific initiatives, and potentially reducing their political support. | *Reprogram unobligated funds:* Redirect IRA funds that have not yet been obligated to other areas, as allowed within administrative limits, to dilute the focus on EV mandates or other climate and clean energy initiatives. | *Defund specific programs through the budget process:* Explicitly cut funding for IRA programs, such as the EV mandates or home energy rebates, during the appropriations process, a strategy that requires coordination with Congress. |
| *Delaying appointments of key personnel:* Delay the appointment of officials responsible for IRA implementation, causing bottlenecks in program execution. | *Commission Reviews of IRA Projects:* Initiate additional review processes or investigations for IRA-supported projects, delaying or deterring implementation. | *Reverse administrative rules established under the IRA:* Undertake a rulemaking process to revoke or replace rules established under the IRA, which is time-intensive and open to legal challenges. |

In parallel, the long-term impact of these policies is poised to exert considerable pressure on both the supply and demand dynamics of critical minerals and rare earth elements. As the IRA's battery and component requirements become more rigorous each year, demand for minerals such as lithium, cobalt, nickel, and various rare earth elements is expected to surge. This increased demand is driving investments in domestic extraction, processing, and refining operations, as manufacturers seek to comply with the new sourcing criteria. However, this shift also raises potential challenges. The accelerated push for domestic production may strain existing mineral reserves and infrastructure, leading to supply bottlenecks and heightened global competition over these resources. Moreover, the race to secure and process critical minerals is intertwined with environmental and social concerns. Communities near mining operations could face environmental degradation and health risks, while pressures to rapidly expand production might compromise labour and sustainability standards. In response, there is a growing impetus for innovation in recycling and extraction technologies, as well as for the establishment of stricter environmental regulations and ethical supply chain practices. Ultimately, while IRA was designed to strengthen the U.S. EV supply chain and reduce foreign dependency, it also catalyses a broader global re-evaluation of resource management, underscoring the need for sustainable practices in the mining and processing sectors amidst mounting long-term demand.

# 3 Pathways to Sustainable Resource Extraction

## 3.1 Ethical Sourcing and Community Integration

A sustainable pathway for resource extraction in the EV industry is not only essential to meet growing demand but also central to addressing the social and environmental costs of conventional mining practices. Ethical sourcing and community integration are fundamental to reimagining extraction, ensuring that affected communities are active stakeholders in the process. Drawing on insights from Latin American studies on green extractivism, ethical sourcing advocates for practices that prioritize local well-being and mitigate exploitative dynamics endemic to current resource frontiers (Fornillo & Lampis, 2023; Lazaro & Serrani, 2023; Svampa, 2019). By involving communities directly in decision-making, companies can foster fair labour standards and ensure cultural and environmental resources are respected. For instance, in Chile's lithium-rich Atacama region, collaboration with Indigenous communities can reduce conflicts over water rights and land use while promoting socio-environmental justice. Ethical sourcing should therefore go beyond mere compliance, becoming an intentional practice that aligns corporate operations with community needs, building a model of extraction that is inclusive, fair, and sustainable (Bebbington et al., 2018).

## 3.2 Responsible Mining Practices

Responsible mining practices are crucial to sustainable resource extraction, ensuring environmental protection and social equity. In the Democratic Republic of Congo, formalizing the cobalt mining sector could significantly enhance safety conditions, environmental stewardship, and the livelihoods of artisanal miners, who often operate in hazardous conditions. However, this shift requires robust corporate responsibility frameworks to hold companies accountable for their labour and environmental impact (Sovacool et al., 2021). Effective responsible mining includes stricter environmental controls, such as monitoring water usage and managing waste to mitigate ecological degradation. Additionally, companies must go beyond superficial sustainability efforts, committing to fair wages, workplace safety, and gender equity within supply chains. As consumer awareness and expectations for ethical practices grow, responsible mining is becoming integral to fostering a resilient EV industry that prioritizes both human rights and environmental planet (Hilson, 2016; Hilson & Hu, 2022).

The establishment of responsible supply chains has been formalized through initiatives such as the Aluminium Stewardship Initiative, Responsible Cobalt Initiative, The Copper Mark, and ResponsibleSteel™ (Giurco, 2020). While these multi-stakeholder processes require time to effect meaningful change, they are evolving to set more ambitious standards while ensuring broad industry participation. Sustainable procurement, driven by both corporations and governments, is playing an increasing role in promoting ethical sourcing. Minerals and metals form the backbone of the global economy and are critical to achieving the United Nations Sustainable Development Goals (SDGs) and the Paris Agreement (Parra et al., 2020). Consequently, mineral resource governance has taken a central role in economic development agenda at the national, regional and international levels, with growing recognition that mining must continue to meet future demand (IEA, 2023). While mining in ecologically sensitive areas raises concerns, the international community must plan for a sustainable future where responsible mining supports long-term economic and environmental objectives, aligning with initiatives such as the 10-Year Framework on Responsible Consumption and Production and the One Planet Network. Additionally, extended producer responsibility (EPR) has emerged as a critical mechanism for ensuring sustainable end-of-life management for products containing critical minerals (Grootenhuis et al., 2024; Sovacool et al., 2020). In the European Union, EPR policies mandate that EV battery producers oversee the proper collection and recycling of lithium, cobalt, and nickel, reducing environmental impact and dependence on raw material imports. Similarly, the EU's Waste Electrical and Electronic Equipment (WEEE) Directive holds solar panel producers accountable for collecting and recycling end-of-life panels, leading to the development of effective recycling infrastructures (Valero et al., 2021b). In India, electronic waste regulations now include solar and wind energy equipment, requiring producers to manage end-of-life treatment, though further strategies are needed to enhance profitability and efficiency. In the United States, discussions are emerging around implementing EPR policies for EV batteries (Wu & Moerenhout, 2024), aiming to hold manufacturers responsible for product lifecycles, encouraging sustainable practices, and improving recycling efficiency.

## 3.3 Sustainable Recycling and Circular Economies

The circular economy aims to achieve sustainable economic growth by maintaining the value of products, materials, and resources for as long as possible, thereby minimizing waste and reducing environmental pressures, particularly on mining (Wu et al., 2024; Valero et al., 2021a). Rooted in economic theories related to industrial metabolism and material flows, this model shifts from the traditional linear system of "produce-use-dispose" to a regenerative approach that promotes reuse and recycling. A key application of circular economy principles is lithium-ion battery recycling, which provides a secondary supply of critical materials such as lithium, cobalt, and nickel, lessening dependency on primary resource extraction and mitigating mining-related environmental impacts. For instance, global implementation of the European Union's battery recycling regulations could reduce primary lithium demand by over 50%, highlighting the role of recycling in addressing resource limitations and enhancing sustainability (Carrasco et al., 2023; Johnson et al., 2024; Luong et al., 2022). In automotive manufacturing, studies assessing the criticality of components have identified 31 essential parts among 794 examined, classified based on rarity and intensity values. However, batteries fall outside this classification due to their inclusion under Directive 2006/66/EC, which mandates dedicated recycling processes at designated treatment centres to mitigate environmental risks effectively (Bown, 2023; Martins & Castro, 2020, 2020; Ortego et al., 2018). By integrating circular economy strategies, industries can adopt a more selective and efficient approach to resource recovery, as demonstrated by thermodynamic criticality assessments that guide sustainability efforts across automotive materials. Tables 1.3 and 1.4 summarize the thermodynamic criticality of these components across different automotive materials.

Investing in recycling technologies and infrastructure for battery materials supports the development of a sustainable supply chain, one that meets production needs while minimizing the environmental degradation tied to raw material extraction. This approach aligns with a broader strategy for advancing resource efficiency, reducing waste, and establishing a more resilient production ecosystem, all of which are essential to supporting the transition to a circular economy. By integrating these practices, companies can lessen their reliance on finite resources, contribute to environmental protection, and ensure a reliable supply of essential materials.

Table 1.3 Number of vehicle parts in each category (Ortego et al., 2018)

| Category | Number of components |
|---|---|
| Category > 1GJ | 2 |
| Category A | 0 |
| Category B | 0 |
| Category C | 0 |
| Category D | 2 |
| Category E | 4 |
| Category F | 7 |
| Category G | 16 |
| Category H | 49 |
| Category I | 714 |
| Critical components | 31 |
| Total components | 794 |

Table 1.4 List of 31 critical components identified in the vehicle (Ortego et al., 2018)

| Critical Components | Rarity (kJ) | Intensity (kJ/kg) | Category |
|---|---|---|---|
| Engine | 32,099,655 | 191.52 | |
| Gearbox | 1,734,110 | 33.14 | |
| Infotainment unit | 538,665 | 389.19 | D |
| On-board supply control unit | 250,164 | 677.9 | D |
| Front axle | 849,407 | 26.81 | E |
| Exhaust gas temperature sensor | 44,319 | 785.39 | E |
| Aerial amplifier (left) | 35,217 | 845.45 | E |
| Aerial amplifier (right) | 35,217 | 845.45 | E |
| Battery | 518,856 | 25.19 | F |
| Combi instrument | 269,510 | 367.46 | F |
| Airbag control unit | 113,557 | 529.44 | F |
| Door control unit (driver) | 60,603 | 610.61 | F |
| Door control unit (passenger) | 59,165 | 600.97 | F |
| Lamp for ambience lighting (driver) | 2,148 | 511.3 | F |
| Lamp for ambience lighting (passenger) | 2,148 | 511.3 | F |
| Generator | 453,144 | 71.54 | G |
| Intermediate exhaust pipe with rear silencer | 450,584 | 32.45 | G |
| Starter | 344,093 | 82.64 | G |
| Wiring | 326,124 | 172.64 | G |
| Rear lighting wiring | 265,318 | 202.3 | G |
| Exterior rear mirror (left) | 209,219 | 180.97 | G |
| Exterior rear mirror (right.) | 202,880 | 230.21 | G |
| Front lighting engine wiring | 188,196 | 188.48 | G |
| Rear screen cleaner motor | 182,001 | 230.49 | G |
| Additional brake light | 31,345 | 354.58 | G |
| Lighting switcher | 26,871 | 362.84 | G |

| Rain sensor | 7,124 | 431.08 | G |
|---|---|---|---|
| Air quality sensor | 4,922 | 346.8 | G |
| Speed sensor (front left tire) | 4,548 | 450.91 | G |
| Speed sensor (front right tire) | 4,548 | 450.91 | G |
| Cable shoe used for anti-twist device | 1,727 | 352.43 | G |

Beyond environmental benefits, sustainable recycling practices align closely with the ethical imperative of reducing dependence on conflict-prone extraction zones. The recycling of materials used in EV batteries reduces the industry's reliance on the DRC and other regions where mining practices often exploit vulnerable labour forces and damage ecosystems (Luong et al., 2022; Mayyas et al., 2023; Zhang et al., 2023). Moreover, circular economies create opportunities for local economies by generating jobs in the recycling and remanufacturing sectors, building a supply chain that is both economically viable and ethically responsible. As research and development advance in this field, the potential for closed-loop systems, where materials are continually reused, offers a vision for a more resilient and ethical industry. Such innovations demonstrate that achieving sustainability is not limited to reducing harm at the extraction phase but includes optimizing every stage of the material lifecycle, ultimately advancing a greener, fairer supply chain.

The convergence of ethical sourcing, responsible mining, and sustainable recycling forms a comprehensive framework that can guide the future of EV resource extraction. Integrating these practices into a cohesive strategy requires commitment from all stakeholders—governments, corporations, and communities—to uphold standards that promote environmental protection and social justice. This framework encourages not only a shift in business operations but also a cultural transformation within the industry, wherein long-term sustainability is prioritized over immediate gains (Schwab & Combariza Diaz, 2023). Policymakers can strengthen these initiatives by mandating transparent reporting on environmental and social impacts, while consumers can drive accountability by supporting brands that demonstrate genuine commitment to sustainable practices. As the world transitions to electrified transportation, the path forward must be informed by a nuanced understanding of the ethical and environmental stakes involved in resource extraction. Only through such a multifaceted approach can the EV industry ensure that the promise of sustainable transportation is not achieved at the expense of communities or ecosystems but rather becomes a driver of global well-being.

# 4 Towards Equitable and Sustainable Electrified Transportation

## 4.1 Collaborative Efforts for Sustainability

Achieving an equitable and sustainable future for electrified transportation requires a holistic approach that aligns environmental sustainability with social justice. The rapid expansion of EV markets, while instrumental in reducing greenhouse gas emissions, has simultaneously escalated demand for critical raw materials, posing ethical and environmental challenges. These challenges disproportionately impact resource-rich, economically vulnerable communities, often creating significant socio-environmental costs that are neither acknowledged nor mitigated by those benefiting from the shift to electrification. Therefore, an ethical EV supply chain must go beyond emissions reduction, incorporating principles of fair labour, community engagement, and ecological preservation (Bebbington et al., 2018). Establishing an inclusive model of electrification demands coordinated efforts from governments, corporations, and local communities. Through cross-stakeholder engagement, sustainability initiatives can be tailored to meet local needs, ensuring that communities central to resource extraction play a participatory role in the EV supply chain and its impact on their environment and livelihoods.

Central to this collaborative framework is the integration of local knowledge and priorities into sustainability initiatives. Involving affected communities in the decision-making process respects their unique perspectives and allows them to have a say in how their lands and resources are utilized. For example, in the lithium-rich regions of the Andes, Indigenous communities have long maintained a delicate balance with the land that modern mining disrupts. These communities possess invaluable knowledge about the ecological dynamics of their environments, which, if integrated into mining and recycling practices, could minimize ecological harm (Gudynas, 2016). Transparent dialogue between corporations and local stakeholders, supported by government oversight, can foster responsible sourcing practices that protect water resources, reduce land degradation, and enhance labour rights. Such inclusive approaches are not only ethically sound but also contribute to long-term economic stability by ensuring that local ecosystems and economies remain resilient in the face of industrial change. When community voices guide sustainability initiatives, they help craft a resource extraction model that values both immediate and future well-being, thus providing a foundation for a more equitable electrification strategy.

### 4.2 Innovative Solutions and Future Directions

A sustainable electrified transportation future also depends on innovations that reduce dependency on traditional, high-impact extraction methods. Future research must prioritize scalable, equitable solutions in resource management, including advancements in alternative materials, which could alleviate the pressure on heavily mined resources like lithium and cobalt. Developing materials that are abundant, recyclable, or require less energy-intensive extraction processes could redefine the resource needs of EV production, fostering a more resilient supply chain. Additionally, researchers are exploring bio-based and synthetic alternatives that could one day replace minerals in EV batteries, reducing dependency on conflict-prone regions and ecologically sensitive areas (Banza Lubaba Nkulu et al., 2018). By investing in such research, the EV industry can minimize its environmental footprint while protecting vulnerable communities from the adverse effects of resource extraction. The adoption of alternative materials offers a promising direction, both mitigating ecological degradation and advancing more ethical, inclusive models of technological innovation that support the transition to electrified transport.

The creation of a circular economy further complements efforts toward a sustainable EV future. Recycling infrastructure, particularly for lithium-ion batteries, is essential to closing the loop in the supply chain, enabling a shift from a linear to a circular model. This approach minimizes the need for raw material extraction by reusing materials from spent batteries, thereby reducing waste and alleviating the environmental impact associated with mining activities (Anlauf, 2017; Carrasco et al., 2023). Expanding global recycling capabilities could notably decrease reliance on newly mined materials, potentially halving the demand for primary lithium if adopted on a large scale (Ortego et al., 2018). As battery recycling technology advances, it holds the potential to create local employment opportunities in recycling plants and remanufacturing facilities, promoting economic growth within communities previously reliant on extractive industries. By prioritizing circular economy models, the EV sector can significantly lower the environmental costs of battery production, making the transition to electrified transportation more sustainable and less exploitative for resource-dependent communities.

Innovation in decentralized energy storage represents another avenue for promoting sustainable electrification. Decentralized storage solutions, such as community-based energy storage systems, enable energy independence by reducing reliance on large-scale grid infrastructure and centralized power sources. This approach is particularly valuable in remote areas and developing regions, where grid access is limited, and energy demands are often met through unsustainable practices. Community-owned energy storage can provide localities with greater control over their energy sources, contributing to a more resilient and self-sustaining energy landscape that aligns with the ethos of environmental justice (Y. Wu et al., 2024). Decentralized storage technologies also mitigate transmission losses and promote efficient energy distribution, reducing the overall demand for high-capacity, grid-reliant systems. By empowering communities to produce and store energy locally, decentralized storage solutions lay the groundwork for a sustainable electrification framework that serves both urban and rural populations, fostering equity in the transition to renewable energy sources.

Ultimately, the vision for equitable and sustainable electrified transportation requires a convergence of technological innovation, community empowerment, and policy commitment. Governments play a critical role in enforcing policies that mandate ethical sourcing, environmental safeguards, and transparent supply chain practices. Corporations must go beyond compliance, committing to genuine social responsibility through transparent reporting, fair labour practices, and investment in sustainable technologies. Consumers, too, have an influential role, as informed purchasing decisions and demand for transparency drive corporate accountability (Schwab & Combariza Diaz, 2023). When these efforts are unified within a cohesive framework, they enable a systemic transformation that transcends traditional boundaries between environmental and social objectives. This holistic approach can help create an electrified transportation system that not only addresses the climate crisis but also respects human rights, upholds environmental integrity, and builds resilient communities (Nyangon, 2024). Through shared responsibility and innovative collaboration, the path toward a truly sustainable, equitable EV industry is within reach, offering a blueprint for industries aiming to balance growth with ethical stewardship and long-term sustainability.

# 5 Conclusion

In concluding, the challenges inherent in resource extraction for EV production underscore a pressing need for transformation in how materials are sourced, mined, and recycled. The transition to an electrified transportation future, essential for meeting global climate objectives, cannot come at the cost of environmental degradation, social exploitation, or inequity (Byrne et al., 2022). Embracing ethical sourcing, responsible mining, and a circular economy for battery recycling is critical. These strategies must collectively drive change across the entire EV supply chain, ensuring that electrification advances not only in a manner that reduces emissions but also respects the rights and well-being of communities in resource-rich areas (Svampa, 2019). IRA in the United States exemplifies regulatory efforts that can catalyse this transformation. By mandating responsible sourcing and prioritizing human rights and environmental protections, the IRA sets a new standard, incentivizing companies to adopt sustainable practices throughout their supply chains. However, its effectiveness will ultimately depend on its ability to be adapted to global supply networks, underscoring the importance of international cooperation and unified policy frameworks in realizing a truly sustainable and equitable future for EV production (Bebbington et al., 2018).

Under the Second Trump Administration, the implementation and enforcement of IRA are poised to undergo significant alterations, reflecting an "America First" approach. Regulatory agencies, guided by administration priorities, could pause funding rollouts, revise rules, and limit loan approvals for critical initiatives such as EV tax credits and hydrogen production. Such adjustments might weaken strict domestic sourcing criteria for battery components, promoting imports from traditional allies over fostering robust domestic manufacturing. While this could alleviate immediate supply chain constraints and align with deregulatory aims, it risks undermining incentives for U.S.-based production and long-term supply security. These shifts may also cause policy uncertainty, deterring large-scale investments in the EV supply chain and potentially compromising the global competitiveness of U.S. industries. Furthermore, federal funding for finalized projects might face redistribution toward traditional energy priorities, emphasizing fossil fuels and deregulation.

Addressing the implications of these policy shifts demands a polycentric strategy (Byrne et al., 2022; Nyangon, 2017, 2021). Policymakers must actively engage local stakeholders in resource management decisions, ensuring equity and environmental stewardship are prioritized alongside industrial development. By aligning short-term policy objectives with the broader goals of sustainability and justice, the EV industry can emerge as a transformative force that reconciles the complexities of ethical resource use with the imperatives of electrification. Ensuring stable incentives for domestic mineral extraction, processing, and refining is crucial to mitigate reliance on volatile global markets for critical resources like lithium and cobalt. This approach would position the U.S. as a leader in sustainable transportation, bridging the tension between market dynamics and global climate commitments while safeguarding long-term supply security.

**List of Abbreviations**

BIPOC Black, Indigenous, and People of Color
DOE U.S. Department of Energy
DRC Democratic Republic of the Congo
EPR Extended Producer Responsibility
EV Electric Vehicles
FDI Foreign Direct Investment
GGRF DOE’s Loan Programs Office or the EPA’s Greenhouse Gas Reduction Fund
IEA International Energy Agency
IRA U.S. Inflation Reduction Act
LCA Life cycle assessment
SDG United Nations Sustainable Development Goals
WEEE EU’s Waste Electrical and Electronic Equipment Directive

**Appendix**

Appendix A Critical minerals and rare earth elements essential for the energy transition and related activities

| Mineral | Applications | Significance | Representative Countries | Potential Injustices |
|---|---|---|---|---|
| Aluminium | Lightweight, transportation, packaging, construction | Energy-efficient, recyclable, infrastructure, low density | China, Russia, Canada, Australia, India | Environmental degradation from bauxite mining, worker exploitation, pollution from refining, land displacement |
| Antimony | Flame retardants, batteries, alloys, electronics | Energy storage, flame resistant, metal alloying, technological applications | China, Russia, Bolivia, Tajikistan, South Africa | Toxicity exposure, labour rights issues, environmental pollution, resource conflicts |
| Arsenic | Semiconductors, glass, pesticides, wood preservatives | Semiconductor materials, industrial uses, pesticide alternative, material processing | China, Russia, USA, Canada, Argentina | Health hazards from exposure, environmental contamination, occupational safety, disposal issues |
| Baryte | Oil drilling, paints, rubber, pharmaceuticals | Drilling fluid density, industrial applications, chemical processing, medical uses | China, India, Morocco, Mexico, USA | Health risks for workers, environmental impact from mining, chemical exposure, waste management |
| Beryllium | Aerospace, electronics, nuclear reactors, alloys | High strength, lightweight, heat resistant, critical for aerospace | USA, China, Russia, Canada, Brazil | Chronic beryllium disease, worker exposure risks, toxic dust, limited recycling |
| Bismuth | Alloys, pharmaceuticals, cosmetics, fire detectors | Non-toxic alternatives, medical applications, cosmetic uses, safety devices | China, Mexico, Bolivia, Argentina, Peru | Mining impacts, chemical exposure, limited recycling, environmental pollution |
| Caesium & Rubidium | Electronics, oil drilling, research, specialty glasses | High-tech applications, drilling operations, scientific research, material manufacturing | Canada, Kazakhstan, Russia, Zimbabwe, Brazil | Resource scarcity, mining labour issues, environmental contamination, limited recycling |
| Chromium | Stainless steel, plating, refractory materials, catalysts | Corrosion resistance, industrial strength, high melting point, catalysis | South Africa, Kazakhstan, India, Turkey, Russia | Toxic emissions from plating, mining pollution, worker health hazards, waste disposal |
| Cobalt | Batteries, superalloys, catalysts, electronics | Energy storage, high-performance materials, catalytic processes, technology integration | Democratic Republic of Congo, Russia, Australia, Philippines, Cuba | Child labour, conflict financing, environmental degradation, worker exploitation |
| Copper | Electrical wiring, plumbing, renewable | Conductivity, essential infrastructure, renewable energy systems, durable material | Chile, Peru, China, USA, Congo | Mining pollution, labour rights violations, environmental damage, water contamination |
| | energy, electronics | | | |
| Fluorspar | Steelmaking, aluminium production, chemicals, electronics | Flux agent, metallurgical applications, chemical manufacturing, semiconductor industry | China, Mexico, Mongolia, Russia, South Africa | Mining hazards, environmental pollution, worker health risks, toxic waste management |
| Gallium | Semiconductors, LEDs, solar cells, electronics | High-tech devices, energy-efficient lighting, renewable energy, electronics manufacturing | China, Germany, Kazakhstan, USA, India | Resource scarcity, mining environmental impacts, toxic waste, limited recycling |
| Germanium | Fiber optics, solar cells, infrared optics, electronics | Optoelectronics, renewable energy, high-precision optics, semiconductor devices | China, Russia, Canada, United States, Belgium | Environmental impacts, toxic byproducts, mining pollution, recycling challenges |
| Graphite | Batteries, steelmaking, lubricants, refractories | Energy storage, industrial processes, lubrication, high-temperature applications | China, Mozambique, Canada, USA, India | Environmental degradation, worker exposure to graphite dust, mining conflicts, recycling issues |
| Hafnium | Nuclear reactors, aerospace, superalloys, electronics | High melting point, corrosion resistance, critical for nuclear industry, high-tech applications | Australia, Russia, South Africa, USA, Brazil | Mining environmental impact, limited sources, high cost, toxic processing |
| Helium | Cooling systems, MRI machines, space applications, electronics | Essential for cooling, medical imaging, space technology, semiconductor manufacturing | USA, Qatar, Algeria, Russia, Canada | Resource scarcity, high extraction costs, geopolitical dependence, supply disruptions |
| Indium | Touchscreens, solar cells, LEDs, alloys | High-tech displays, renewable energy, energy-efficient lighting, material enhancement | China, South Korea, Canada, Japan, Belgium | Mining environmental impacts, labour rights issues, resource scarcity, recycling challenges |
| Lithium | Batteries, energy storage, electric vehicles, electronics | Energy storage, renewable energy integration, mobility solutions, high demand | Australia, Chile, China, Argentina, USA | Water resource depletion, labour exploitation, geopolitical tensions, biodiversity issues |
| Magnesium | Lightweight alloys, automotive, aerospace, electronics | Weight reduction, high strength, heat resistance, essential for high-tech manufacturing | China, Russia, USA, Israel, Canada | Mining pollution, energy-intensive production, worker safety, recycling limitations |
| Manganese | Steelmaking, batteries, alloy production, chemical applications | Strengthening steel, energy storage, industrial applications, chemical processes | South Africa, Australia, China, Gabon, Brazil | Environmental degradation, worker exploitation, mining conflicts, pollution |
| Nickel | Batteries, stainless steel, | Energy storage, corrosion resistance, high-performance | Indonesia, Philippines, | Mining pollution, labour rights violations, |
| | alloys, electronics | materials, technological applications | Russia, Canada, Australia | environmental harm, resource conflicts |
| Niobium | Superalloys, steelmaking, electronics, aerospace | High-strength materials, industrial infrastructure, high-tech manufacturing, aerospace components | Brazil, Canada, Australia, Nigeria, Russia | Mining impacts, worker exploitation, environmental degradation, limited recycling |
| Platinum Group Metals (PGMs) | Catalysts, automotive, electronics, jewellery | Emission control, high-tech devices, industrial processes, luxury goods | South Africa, Russia, Zimbabwe, Canada, USA | Mining environmental damage, labour rights issues, resource conflicts, recycling challenges |
| Potash | Fertilizers, agriculture, industrial chemicals, food production | Essential for agriculture, food security, industrial applications, economic importance | Canada, Russia, Belarus, China, India | Environmental impacts from mining, water usage conflicts, land degradation, labour issues |
| Rare Earth Elements | Magnets, electronics, renewable energy, defence | High-tech applications, energy transition, military technology, industrial uses | China, USA, Australia, Russia, India | Environmental pollution, labour exploitation, geopolitical tensions, resource conflicts |
| Rhenium | Superalloys, catalysts, electronics, petrochemical processes | High-temperature resistance, catalytic processes, high-tech manufacturing, industrial applications | Chile, USA, Kazakhstan, Portugal, Germany | Mining environmental impact, resource scarcity, worker safety, recycling limitations |
| Scandium | Aerospace, electronics, lighting, sports equipment | High-strength materials, high-tech devices, energy-efficient lighting, lightweight manufacturing | China, Russia, Australia, USA, Kazakhstan | Limited supply, high cost, environmental impact of mining, recycling challenges |
| Silver | Electronics, solar cells, photography, jewellery | Conductivity, renewable energy, industrial uses, luxury goods | Mexico, Peru, China, Russia, Australia | Mining pollution, labour rights, environmental degradation, resource conflicts |
| Strontium | Magnets, fireworks, electronics, ferrite magnets | High-temperature applications, pyrotechnics, electronic components, magnetic materials | China, Mexico, India, Spain, USA | Environmental contamination, mining pollution, worker health risks, recycling issues |
| Tantalum | Electronics, aerospace, medical devices, alloys | Capacitors, high-performance materials, medical applications, industrial uses | Democratic Republic of Congo, Rwanda, Brazil, Australia, Canada | Conflict minerals, child labour, environmental degradation, unethical sourcing |
| Tellurium | Solar cells, electronics, alloys, thermoelectric | Renewable energy, high-tech devices, material enhancement, energy conversion | China, Canada, Peru, Bolivia, USA | Mining pollution, resource scarcity, toxic waste, recycling challenges |
| Tin | Soldering, electronics, packaging, alloys | Electrical connections, high-tech manufacturing, | China, Indonesia, Peru, Bolivia, Myanmar | Mining environmental impacts, child labour, resource conflicts, pollution |
| | | packaging materials, metal alloying | | |
| Titanium | Aerospace, medical implants, pigments, alloys | High strength, biocompatible, corrosion resistant, lightweight materials | China, Russia, Japan, USA, Kazakhstan | Mining impacts, energy-intensive processing, worker safety, environmental degradation |
| Tungsten | Electronics, steelmaking, mining machinery, lighting | High melting point, Hardness, Industrial tools, Lighting applications | China, Russia, Vietnam, Austria, Portugal | Mining pollution, worker exploitation, environmental degradation, resource conflicts |
| Uranium | Nuclear energy, medical isotopes, military applications, research | Energy production, medical uses, national security, scientific research | Kazakhstan, Canada, Australia, Namibia, Russia | Radioactive contamination, worker exposure, environmental impact, nuclear waste disposal |
| Vanadium | Steel alloys, batteries, catalysts, aerospace | Strengthening steel, energy storage, industrial catalysis, high-performance materials | China, Russia, South Africa, Australia, USA | Mining environmental impacts, worker health risks, resource scarcity, pollution |
| Zirconium & Hafnium | Nuclear reactors, Aerospace, Electronics, Ceramics | Nuclear applications, high-tech manufacturing, corrosion resistance, specialized ceramics | Australia, South Africa, China, USA, Russia | Mining environmental impacts, limited sources, worker safety, recycling challenges |

Sources: Chen et al., 2024; Nyangon & Darekar, 2024; Agusdinata & Liu, 2023; IEA, 2023; IMF, 2023; Forget & Bos, 2022; Dall-Orsoletta et al., 2022; Valero et al., 2021c; Nyangon & Byrne, 2021; Habib et al., 2020; Parra et al., 2020; Banza Lubaba Nkulu et al., 2018; Diemer et al., 2018; Nyangon et al., 2017; Sovacool, 2017; Sovacool & Dworkin, 2014